\documentstyle[12pt]{article}
\newcommand{\newc}{\newcommand}
\newc{\renewc}{\renewcommand}
\newc{\pageyear}[2]{, #1 (19#2)}
\newc{\yearpage}[2]{ (19#2), #1}
\newc{\pandy}[2]{\yearpage{#1}{#2}}
\newc{\ibid}[3]{{\em ibid.\/}\ {\bf #1}\pandy{#2}{#3}}
\newc{\pl}[3]{Phys.\ Lett.\ {\bf #1}\pandy{#2}{#3}}
\newc{\plb}[3]{Phys.\ Lett.\ B\ {\bf #1}\pandy{#2}{#3}}
\newc{\prl}[3]{Phys.\ Rev.\ Lett.\ {\bf #1}\pandy{#2}{#3}}
\newc{\np}[3]{Nucl.\ Phys.\ {\bf #1}\pandy{#2}{#3}}
\newc{\npb}[3]{Nucl.\ Phys.\ {\bf B#1}\pandy{#2}{#3}}
\newc{\rpp}[3]{Rep.\ Prog.\ Phys.\ {\bf #1}\pandy{#2}{#3}}
\newc{\yf}[3]{Yad.\ Fiz.\ {\bf #1}\pandy{#2}{#3}}
\newc{\sjnp}[3]{Sov.\ J.\ Nucl.\ Phys.\ {\bf #1}\pandy{#2}{#3}}
\newc{\zph}[3]{Z.\ Phys.\ {\bf #1}\pandy{#2}{#3}}
\newc{\ijmpa}[3]{Int.\ J.\ Mod.\ Phys.\ A {\bf #1}\pandy{#2}{#3}}
\newc{\prd}[3]{Phys.\ Rev.\ D {\bf #1}\pandy{#2}{#3}}
\newc{\zetp}[3]{Zh.\ Eksp.\ Teor.\ Fiz.\  {\bf #1}\pandy{#2}{#3}}
\newc{\jetp}[3]{Sov.\ Phys.\ JETP {\bf #1}\pandy{#2}{#3}}
\newc{\jmp}[3]{J.\ Math.\ Phys.\  {\bf #1}\pandy{#2}{#3}}
\newc{\bm}[1]{\mbox{\boldmath $#1$}}
\newc{\bvec}{\bm{B}}
\newc{\hvec}{\bm{H}}
\newc{\evec}{\bm{e}_3}
\newc{\ezvec}{\bm{e}_3}
\newc{\exvec}{\bm{e}_1}
\newc{\eyvec}{\bm{e}_2}
\newc{\rhovec}{\bm{\rho}}
\newc{\tauvec}{\bm{\tau}}
\newc{\avec}{\bm{A}}
\newc{\abvec}{\overline{\bm{A}}}
\newc{\zvec}{\bm{Z}}
\newc{\yvec}{\bm{Y}}
\newc{\xvec}{\bm{X}}
\newc{\nvec}{\bm{n}}
\newc{\svec}{\bm{s}}
\newc{\xivec}{\bm{\xi}}
\newc{\rvec}{\bm{r}}
\newc{\txt}{\textstyle}
\newc{\dsp}{\displaystyle}
\newc{\scr}{\scriptstyle}
\newc{\teeny}{\scriptscriptstyle}
\newc{\be}[1]
{\begin{equation} \mbox{$\label{#1}$}}
\newc{\bea}[1]
{\begin{eqnarray} \mbox{$\label{#1}$}\txt}
\newc{\ee}{\end{equation}}
\newc{\eea}{\end{eqnarray}}
\newc{\non}{\nonumber}
\newc{\newl}{\non\\*}
\newc{\beanon}{\begin{eqnarray*}}
\newc{\eeanon}{\end{eqnarray*}}
\newc{\eref}[1]{equation\,(\ref{#1})}
\newc{\pref}[1]{(\ref{#1})}
\newc{\ie}{i.\frenchspacing e.\nonfrenchspacing \ }
\newc{\eg}{e.\frenchspacing g.\nonfrenchspacing \ }
\newc{\cc}{\mbox{$c.c.$ }}
\newc{\cf}{cf.\ }
\newc{\comma}{\hspace{.15cm} {\rm ,}\hspace{-.15cm}}
\newc{\period}{\hspace{.15cm} {\rm .}\hspace{-.15cm}}
\newc{\semic}{\hspace{.15cm} {\rm ;}\hspace{-.15cm}}
\newc{\eps}{\epsilon}
\newc{\epsc}{\epsilon^{\ast}}
\newc{\lam}{\lambda}
\newc{\lamc}{\lambda^{\ast}}
\newc{\muc}{\mu^{\ast}}
\newc{\rp}{\mbox{$r${\small'}}}
\newc{\half}{{{1}\over{2}}}
\newc{\quarter}{{{1}\over{4}}}
\newc{\third}{{{1}\over{3}}}
\newc{\rec}[1]{{{1}\over{#1}}}
\newc{\rsqrt}[1]{\mbox{$\displaystyle {{1}\over{\sqrt{#1}}}$}}
\newc{\im}{{\rm Im}~}
\newc{\re}{{\rm Re}~}
\newc{\zc}{z^\ast}
\newc{\rtwo}{{\cal R}^2}
\newc{\rthree}{{\cal R}^3}
\def\bigone{\relax{1\kern-.14cm 1}}
\newc{\swap}{\leftrightarrow}
\newc{\gtapprox}
{{\,}^{\raisebox{-.4ex}{${\scriptstyle>}$}}_
{\raisebox{.4ex}{${\scriptstyle\sim}$}}\,}
\newc{\ltapprox}
{{\,}^{\raisebox{-.4ex}{${\scriptstyle<}$}}_
{\raisebox{.4ex}{${\scriptstyle\sim}$}}\,}

\newc{\sign}[1]{\mbox{${\rm sign}(#1)$}}
\newc{\del}{\mbox{$\bm{\nabla}$}}
\newc{\dv}[1]{\mbox{$d^{\,2}\!#1$}}
\newc{\order}[1]{{\cal O}(#1)}
\newc{\dd}[2]{{{\partial #1}\over{\partial #2}}}
\newc{\ddd}[2]{{{{\partial}^2 #1}\over{\partial {#2}^2}}}
\newc{\dddd}[3]{{{{\partial}^2 #1}\over
{\partial #2 \partial #3}}}
\newc{\smalldag}{{\mbox{{\footnotesize\dag}}}}
\newc{\bra}[1]{\langle#1|}
\newc{\ket}[1]{|#1\rangle}
\newc{\ftensor}[3]{\mbox{$\partial_{#2} #1_{#3} -
\partial_{#3} #1_{#2}$}}
\newc{\ev}[1]{\langle #1 \rangle}
\newtheorem{theorem}{Theorem}
\newtheorem{lemma}{Lemma}
\renewc{\theequation}{\thesection.\arabic{equation}} 
\def\hf {{{1}\over{2}}}
\def\rthf {{{1}\over{\sqrt2}}}
\def\under#1{$\underline {\hbox {#1}}$}
\def\nr {{{n}\over{r}}}
\def\mr {{{m}\over{2r}}}
\def\yr {{{y}\over{r^2}}}

\def\Zno {{${\cal Z}_{\rm NO}\,$}}
\def\zno {{${\rm Z}_{\rm NO}\,$}}
\def\xno {x_{{\rm\teeny NO}}}
\def\yno {y_{{\rm\teeny NO}}}
\def\ea  {{\cal E}_1}
\def\eb  {{\cal E}_2}
\def\ec  {{\cal E}_3}
\def\eno {{\cal E}_0}
\def\ybar{\overline{y}}
\renewc{\thefootnote}{\fnsymbol{footnote}}
\begin{document}
\begin{flushright}
NORDITA -- 94/52 P\\
hep-ph/9410276
\end{flushright}
\vspace*{1cm}
\noindent
{\Large\bf Instability of a Nielsen-Olesen Vortex\\
Embedded in the Electroweak Theory:\footnote
{Talk Presented at the G\"ursey Memorial Conference I on
Strings and Symmetries, Bo\u{g}azi\c{c}i University, \.{I}stanbul, Turkey
6--10 June 1994. To appear in the proceedings
(Springer-Verlag).}\vspace*{1mm}\\
I. The Single-Component Higgs Gauge}

\vspace*{1cm}\noindent
Samuel W. MacDowell$^1$ and
Ola T\"{o}rnkvist$^2$

\vspace*{5mm}\noindent
$^1$~{\small Yale University, Sloane Physics
Laboratory, New Haven, CT 06520, USA}\\
$^2$~{\small NORDITA, Blegdamsvej 17,
DK-2100 Copenhagen \O, Denmark}

\renewc{\thefootnote}{\arabic{footnote}}
\footnotetext[1]{{\tt macdowell@yalph2.physics.yale.edu}}
\footnotetext[2]{{\tt tornkvist@nbivax.nbi.dk}}
\vspace*{1cm}\noindent
{\bf Abstract}. The stability
of an abelian (Nielsen-Olesen) vortex embedded in the electroweak theory
against W production
is investigated in a gauge defined by the condition of a
single-component Higgs field. The model is characterized by the parameters
$\beta=({M_H\over M_Z})^2$ and $\gamma=\cos^2\theta_{\rm w}$  where
$\theta_{\rm w}$ is the weak mixing angle.
It is shown that the equations for W's in the
background of the Nielsen-Olesen vortex have no solutions in the linear
approximation.
A necessary condition for the nonlinear equations to have a solution
in the region of parameter space where the abelian vortex is
classically unstable is that the W's be produced in a state of angular
momentum $m$ such
that $0>m>-2n$. The integer $n$ is defined by the phase of the Higgs field,
$\exp(in\varphi)$. It is shown that, in the region of parameter space
$(\beta, \gamma)$ where the
nonlinear equations have a solution with energy lower than that of the
abelian vortex, this vortex is a saddle point of the energy in the space of
classical field configurations.  Solutions for a set of values
of the parameters $\beta$ and $\gamma$ in this region
were obtained numerically for the case $-m=n=1$. The
possibility of existence of a stationary state for $n=1$ with W's in the
state $m=-1$ was investigated. The boundary conditions for the
Euler-Lagrange equations required to make the energy finite cannot be
satisfied at $r=0$. For these values of $n$ and $m$ the possibility of a
finite-energy stationary state defined in terms of distributions is
discussed.
\newpage
\section*{Introduction}
It has been shown that the Nielsen-Olesen abelian vortex \cite{NieO73}
can be embedded \cite{Nam77,Vac92}
in the electroweak ${\rm SU(2)}\times {\rm U(1)}$ gauge
theory \cite{Wein67,Sal68} in the form of an azimuthal Z field
$Z_{\varphi}(\rho)$ and a lower component of the Higgs field
$\Phi_2=\Phi(\rho)\exp(in\varphi)$, where
$(\rho,\varphi)$ are polar coordinates of the position vector $\rhovec$
perpendicular to the vortex. The embedded vortex, hereafter denoted the
$Z_{NO}$ vortex, is a tube of confined flux of the
Z field strength, which reaches a high value at the center of the vortex.

It is known from previous works that, in a strong uniform magnetic field,
the electroweak vacuum develops an instability through the interaction of
the magnetic field with the anomalous magnetic moment of the W boson,
leading to the formation of a W condensate \cite{AmbO,MacT92,Torn93}.
The magnetic moment interacts similarly with a Z field;
hence an instability with ensuing W production can occur if
the Z field strength is sufficiently high within a region large enough
compared to the Compton wavelength of the W boson.
A measure of these conditions is provided, respectively, by the two parameters
$\beta = (M_H/M_Z)^2$ and $\gamma=(M_W/M_Z)^2 \equiv \cos^2\theta$,
where $M_H,\ M_Z,\ M_W$ are the masses of the Higgs, Z and W bosons
and $\theta$ is the Weinberg mixing angle.
One finds qualitatively that the possibility of instability
increases with higher $\beta$ and higher $\gamma$.

A quantitative investigation of the stability of the \zno vortex
for $n=1$ was performed by James, Perivolaropoulos and Vachaspati
\cite{JamPV}.
They found numerically that the solution becomes
unstable beyond a certain line
in the parameter space $(\beta,\gamma)$. Their analysis was
supplemented with an elegant analytical estimate by Perkins \cite{Perk93},
according to which the \zno solution is unconditionally
unstable for $\gamma>.19$.
In particular, the points $(\beta,\gamma)$
corresponding to the physical value of
the Weinberg angle are inside the region of instability.

In this report we have investigated the problem in a gauge which
maintains the simple structure of the \zno vortex,
defined by the condition that the upper
component of the Higgs field vanishes, rather than the gauge used by
James et al.\ [{\it op.\ cit.\/}] and Ach\'{u}carro et al.\ \cite{AchGHK93}
which allows for a two-component Higgs field in the presence of W bosons.
These two gauges are actually inequivalent since (for $n=1$)
the gauge invariant quantity
$(\Phi_1 ^\smalldag\Phi_1 + \Phi_2 ^\smalldag\Phi_2)$, at $\rho=0$,
is zero in our gauge but non-zero in theirs.

In the first section an ansatz for the fields is presented that preserves
the cylindrical symmetry of the energy density.
The Euler-Lagrange equations are obtained
and the boundary conditions are established.

In Section 2 we consider the equations for W bosons in the background of the
configuration of Higgs and Z fields as given by the \zno vortex.
It is shown that, in this gauge, these equations have no solutions in the
linearized form. On the other hand for particular angular momenta of
the W's, specified by the condition $-2n<m<0$ on the phase $\exp(im\varphi)$
of the polar components of the W field, the set of nonlinear equations may
 admit a solution in a certain domain of the space of
parameters $(\beta,\gamma)$.
 It is shown that, in the region of
parameter space where the nonlinear equations have a solution and the energy,
calculated to lowest order in the W field, is smaller than that for the
\zno vortex, this vortex is not a local minimum but a saddle point
in the space of classical field configurations.

In Section 3, these equations were solved numerically for the case $n=1$,
$m=-1$ and a set of values of the parameters
$\beta=.5,1$ and $\gamma=.25,.5,1$. The energy was computed and found
in each case to be lower than that for the \zno vortex.
As remarked already, the field configurations considered here
are unrelated by any gauge transformation to those considered in
Refs.~\cite{JamPV,AchGHK93}. Hence the instability regions for
the two cases may be different.

The existence of a stationary state with W's,
as suggested in Refs.~\cite{Perk93,Torn92,Oles93},
is considered in Section 4. For $m=-1$, one would expect that there
exists an analytic solution of the
Euler-Lagrange equations for all fields. We found, however, that
the boundary conditions cannot be satisfied at $r=0$.
A discussion is given of the possibility of a vortex state
with W's defined in terms of distributions.

\section{Nonabelian Vortex}

We shall investigate the problem of stability of the \zno vortex
in a gauge fixed by the condition that the Higgs field $\Phi$
has a zero upper component and a lower component
$\Phi(\rho)\exp(in\varphi)$. For the
vortex of the two-dimensional abelian theory, $n$
is a topological winding number defined in terms the total flux of the U(1)
gauge-field strength \cite{NieO73}.
In a non-abelian model $n$ can no longer be defined
in a gauge invariant way. In the above chosen gauge it is given by
$$n={1\over 2\pi i\Phi_0^2}\int (d\Phi^\dagger \wedge d\Phi)$$
in the notation of differential forms,
where $\Phi_0$ is the magnitude of the Higgs field at infinity.
This expression is invariant only under
the electromagnetic U(1) gauge group.

The \zno vortex contains an azimuthal Z field $Z_\varphi$.
One can easily show that, if a radial component depending on the $\rho$
coordinate alone is added to the Z field, the action increases. Therefore,
the vortex solution can only be modified by the inclusion of a W field and
an electromagnetic gauge potential.  The latter can be chosen purely
azimuthal by virtue of the residual electromagnetic gauge invariance.

Let $g,g'$ be the coupling constants for the groups SU(2) and U(1)
respectively. They are related to the Weinberg angle $\theta$ and the
electromagnetic charge $e$ by $g\sin\theta=g'\cos\theta =e$. The physical
gauge fields are related to the gauge potentials ${\bf V}^a$ and ${\bf V}'$
associated with the groups SU(2) and U(1) by
\beanon
{\bf A}&=&{\bf V'}\cos \theta+{\bf V}^3\sin \theta\comma\\*
{\bf Z}&=&{\bf -V'}\sin \theta+{\bf V}^3\cos \theta\comma\\*
{\bf W}&=&{\txt{{1}\over{\sqrt{2}}}}({\bf V}^1-i{\bf V}^2)\period
\eeanon
Let us introduce a dimensionless vector
$${\bf r} =\rhovec\,\Phi_0g/({\sqrt{2}}\cos\theta)\equiv
\rhovec M_Z\comma$$
with polar coordinates $r,\varphi$,
and a set of functions $s,X,Y,Z$ defined by
$$
\Phi=\Phi_{0}s(r)e^{in\varphi}\comma
$$
$$
\begin {array}{lcl}
\rthf V_{\varphi}^{3}\cos\theta=\Phi_{0}Y(r)\comma &\ &
\rthf V_{\varphi}'\sin\theta=\Phi_{0}X(r)\comma \\
\rthf Z_\varphi=\Phi_0(Y-X)=\Phi_0 Z(r)\comma &\ &
\rthf A_\varphi=\Phi_0(Y\tan\theta+X\cot\theta) \period
\end{array}
$$
In order to preserve the vortex cylindrical symmetry the {\bf W} field
must be of the form
$${\bf W}\cos\theta=\Phi_0 [u(r){\bf e}_r+iv(r){\bf e}_\varphi]
\exp(im\varphi)\period$$
It can be shown that the functions $u$ and $v$ may be chosen real without
loss of generality. It is also convenient to use a set of auxiliary
fields
\noindent
\beanon
y&=&Y-{\txt{{m}\over{2r}}}\\*
x&=&X-{\txt{{m}\over{2r}}-{{n}\over{r}}}\\*
z&=&Z+{\txt{{n}\over{r}}}=y-x
\eeanon
and the parameters
$$\beta={\left({\txt{{M_H}\over{M_Z}}}\right)}^2,
\hskip .5cm
\gamma={\left({\txt{{M_W}\over{M_Z}}}\right)}^2=\cos^{2}\theta\period$$

The energy density in terms of these fields and the new variables ${\bf r}$
takes the form
\bea{edens}
{\cal H}&=&\Phi_0^2\,{\left\{(s')^2+((y-x)s)^2
+{{1}\over{4}}\beta (s^2-1)^2 +{1\over \gamma} (v'+{{v}\over{r}}+2yu)^2
\right.}\non\\* ~&+&{\left.{{1}\over{\gamma}}{\Big(}y'+{{y}\over{r}}
-2uv{\Big )}^2  +{{1}\over{1-\gamma}}{\Big (}x'
+{{x}\over{r}}{\Big )}^2+(u^2 +v^2)s^2\right\}}\period
\eea
The vortex energy per unit length is then given by
$\int{\cal H}\,d^2{\bf r}$.
The expression for ${\cal H}$ is invariant under the combined
substitutions $y\to -y$,
$\;x\to -x,\;v\to -v$ (charge conjugation invariance) so it is
sufficient to consider positive values of $n$.

The Euler-Lagrange equations for the fields are
\bea{seq}
&&{\txt s''+{{s'}\over{r}}-[u^2+v^2+(y-x)^2
+{{\beta}\over{2}}(s^2-1)]s=0}\\*
\label{xeq}
&&{\txt x''+{{x'}\over{r}}-{{x}\over{r^2}}+(1-\gamma)(y-x)s^2=0}\\*
\label{yeq0}
&&{\txt y''+{{y'}\over{r}}-{{y}\over{r^2}}-2(2v'u
+vu' +{{1}\over{r}}vu)-4yu^2-\gamma (y-x)s^2=0}\quad\\*
\label{Wphieq}
&&{\txt v''+{{v'}\over{r}}-{{v}\over{r^2}}+2(2y'u+yu'+{{1}\over{r}}yu)
-4vu^2-\gamma v s^2=0}\\*
\label{Wreq}
&&{\txt v'y-vy'+(2y^2+{\gamma\over 2}s^2+2v^2)u=0}\period
\eea
\newpage

We shall also study the equations for W's produced in the background of
the \zno vortex. For this purpose we have to consider the Euler-Lagrange
equations for the fields $u,v,$ and $A_\varphi$, with $s,z$ fixed at their
\zno values. It is convenient to use $y,u,v$ as new independent variables,
so that $x=y-z$. The equations for $u$ and $v$ will be the same,
(\ref{Wphieq}) and (\ref{Wreq}), but the equation for $y$ must be
modified:
\be{yeq}
  y''+{y'\over r}-{y\over r^2}-2(1-\alpha)(2v'u+vu'+{vu\over r}
+2yu^2)-\gamma(y-x)s^2=0,
\ee
where $\alpha=\gamma$ if $A_\varphi$ is allowed to vary or $\alpha=1$
if $A_{\varphi}$ is kept at its \zno value $A_\varphi=0$.
This equation coincides with the previous equation (\ref {yeq0}) if one sets
$\alpha =0$. Therefore the new equation can be used in the three cases with
the appropriate values of $\alpha$. Note that in all cases $\alpha$ is a
non-negative parameter.

{}From the last three equations one obtains the integrability condition
\be{intcond}
{d\over dr}[ru(\gamma s^2+4\alpha v^2)]+2r(\gamma xs^2+4\alpha yu^2)v=0
\period\ee
The full system of equations can be reduced to 4
independent second-order differential equations by solving \pref{Wreq} for
$u$ and \pref{intcond} for $u'$.

  In order to obtain solutions with a finite energy per unit length
of the vortex, the following boundary conditions are imposed.

\vspace*{3ex}
\noindent
\under{Boundary conditions near $r=0$}:
\be{BCr=0}
{\txt s=s_{0}r^{n},\;\;x=-{{2n+m}\over{2r}}+x_{0}r,\;\;
 y=-{{m}\over{2r}}+y_{0}r,\;\; v=v_{0}r^k,\;\; u=u_{0}r^k}
\ee
with $k>-\half$.
Inserting these into the equations, one finds that solutions exist
near the origin in the following three cases.
\be{C1}
\begin{array}{lllc}
{\rm a})& m=0, & k=1,&(1+n)u_0=nv_0\period\\*
{\rm b})& m=k+1, & k=0,1,2\ldots, & u_0=v_0\period\\*
{\rm c})& m=-k-2n-1, & k=0,1,2,\ldots, & mu_0=-(m+2n)v_0\period
\end{array}
\ee
These are the boundary conditions valid for the case $\alpha=0$
corresponding to the variational problem for all the fields or
for the same equations linearized with respect to $u,v$.
For the case of the new equations with $\alpha>0$, the boundary conditions
will still be the same for values of $n,m$ such that $k>n$. On the other hand,
assuming that $k\leq n$, one finds that the equations near $r=0$
can only be satisfied if
\be{BC2}
\begin{array}{lllc}
{\rm d})& m=0, & k=1,\,\,\,n=1,\hspace{.5in} (\gamma s_0^2+4\alpha v_0^2)
u_0={\gamma \over 2}s_0^2v_0\period \\*
{\rm e})& m\neq 0, & k=0,\hspace{1.5in} v_0=mu_0 \period
\end{array}
\ee
We remark however that, for $k=0$, continuity of {\bf W} at the origin
restricts the value of $m$ to be $m=\pm 1$.
\pagebreak[3]

\noindent
\under{Boundary conditions near $r=\infty$}:

\vspace*{2ex}
Depending on the values of $\beta$ and
$\gamma$, different terms in the asymptotic equations are responsible
for the leading exponential behavior at large $r$.
Write $s=1-f$. Assuming that $0<\gamma\leq 1$, we consider first the
simplest case:

\vspace*{2ex}
\noindent i) $\;\beta\leq 4\gamma$, $4\gamma>1$

\vspace*{2ex}
The asymptotic field equations are, to leading order,
\beanon
\mbox{\ref{seq}'}\ : & & {\txt f''+{{f'}\over{r}}-\beta f=0}\\*
\mbox{\ref{xeq}'}\ :& & {\txt z''+{{z'}\over{r}}-{{z}\over{r^2}}-z=0}\\*
\mbox{\ref{Wreq}'}\ :& &
{\txt (v'+{{v}\over{r}}){{y_1}\over{r}} +(2({{y_1}\over{r}})^2+\hf
\gamma)u=0}\\*
\mbox{\ref{yeq}'}\ :& &
{\txt  y''+{{y'}\over{r}}-{{y}\over{r^2}}-\gamma z=0}\\*
\mbox{\ref{intcond}'}\ :& & {\txt u'+{{u}\over{r}}+2y_1 {{v}\over{r}}=0}
\comma
\eeanon
where equation (\ref{xeq}') was obtained by subtracting \pref{xeq}
 from \pref{yeq}.
Differentiating (\ref{intcond}'),
and using (\ref{Wreq}') and (\ref{intcond}') to eliminate $v'$ and
$v$, one obtains an asymptotic equation for $u$,
\be{ueq}
u''+3{{u'}\over{r}}+{{u}\over{r^2}}
-(4({{y_1}\over{r}})^2+\gamma)u=0\period
\ee
{}From equations (\ref{seq}', \ref{xeq}', \ref{yeq}',
\ref{ueq}) and (\ref{intcond}'), one finds the following approximate
solutions for large $r$:
\bea{BCinf}
f(r)&\sim & {\txt f_1 K_0(\sqrt{\beta}\,r)}\non\\*
z(r)&\sim & {\txt z_1 K_1(r)}\non\\*
y(r)&\sim &{\txt {{y_1}\over{r}}+\gamma z}\\*
u(r)&\sim &{\txt -v_1{2y_1\over r}K_{|2y_1|}(\sqrt{\gamma}\,r)}\non\\*
v(r)&\sim &{\txt v_1{{d}\over{dr}}K_{|2y_1|}(\sqrt{\gamma}\,r)\period}\non
\eea
The parameter $y_1$ is related to the total flux of
the electromagnetic field,
$$
\oint_{\infty}\!\! {\bf A}\cdot \bm{d}\rhovec =
{{2\pi}\over{e}}(2 y_1 + 2\gamma n + m)\period
$$

For values of $\beta,\ \gamma$ which do not satisfy condition (i) above,
the expressions for $z$ (or $y$) and $f$ must be modified as follows.

\vspace*{2ex}\noindent
ii) If $4\gamma\leq 1$,

\vspace*{2ex}
\noindent
then the asymptotic expression for $z$ in the set of equations
(\ref {seq}--\ref {Wreq}),
or for $y$ in the set of equations (\ref {Wphieq}--\ref {yeq}),
has to be modified. Making use of the Green function for
\pagebreak[3]

\noindent the Laplace operator
in two dimensions let us define:
\bea{zmod}
\zeta(r) & = & - {{1}\over{2\pi}} \int\!K_0(|{\bf r}-
{\bf r}\mbox{'}|) \cos(\varphi-\varphi\mbox{'})\,j(\rp)\, d^2{\bf r}\mbox{'}
\comma\non\\*
j(r)&\equiv&
2\left({\txt 2v'(r)u(r) + v(r)u'(r) + {{1}\over{r}}v(r)u(r)+
{{2y_1}\over{r}}{[u(r)]}^2} \right).
\eea
Then, for the set of equations (\ref {seq}--\ref {Wreq}), one would have
\be {zas}
z=z_1K_1(r)+\zeta (r)\comma
\ee
while for the set (\ref {Wphieq}--\ref {yeq}) one would have
\be {yas}
y={y_1\over r}+\gamma z_1 K_1(r) +(1-\alpha)\zeta (r)\period
\ee
iii) If $\beta\geq 4\gamma$ or $\beta\geq 4$,

\vspace*{2ex}
\noindent
then the expression for $f$ has to be modified. Similarly to case (ii)
one finds
\be{fmod}
f(r) = {\txt f_1 K_0(\sqrt{\beta}\,r) }+
{{1}\over{2\pi}} \int\!{\txt K_0(\sqrt{\beta}|{\bf r}-
{\bf r}\mbox{'}|)}
\left( z(\rp)^2 + u(\rp)^2 + v(\rp)^2\right)d^2{\bf r}\mbox{'}\period
\ee
In the integrands of (\ref{zmod}, \ref{fmod}), $u$ and $v$ are given by
their asymptotic expressions \pref{BCinf}, and in equation \pref{fmod},
$z$ is given by \pref{BCinf} or \pref{zas} as prescribed by the value of
$\gamma$.

The result \pref{fmod}
means that the leading asymptotic behavior of the Higgs field is
determined,
in case (iii), not by the Higgs mass but by twice the W boson mass or,
in the limit of zero W fields, by
twice the Z mass. This finding agrees with the expressions obtained in
a recent reanalysis of the Nielsen-Olesen vortex \cite{Peri93}.

The asymptotic expressions involve four parameters, $f_1$, $z_1$, $y_1$
and $v_1$.  Together with the four boundary parameters
$s_0$, $x_0$, $y_0$ and $v_0$ at $r=0$, the number of unknowns
equals the rank of the system of differential equations.
Therefore, if a solution to the equations exists, then all parameters
would be determined by imposing the respective boundary conditions at
$r=0$ and at a value $r=r_1\gg 1$.

We shall now investigate the question of existence of a solution with
these boundary conditions.
Let us introduce the functions  $V=v/y$  and $U=ru(\gamma s^2+4\alpha v^2)$.
The equations for $U$ and $V$ are
\bea{Veq}
&&{\txt V'+2(1+{1\over y^2}(v^2+{\gamma\over 4}s^2))u=0}\comma\\*
\label{Ueq}
&&U'+2r(\gamma xs^2+4\alpha yu^2)v=0 \period
\eea
Multiplying \pref{Veq} by $U$, \pref{Ueq} by $V$, and adding
the equations one obtains
\be{UVeq}
{\txt (UV)'+2r}\left\{\left({\txt [1+{1\over y^2}(v^2
+{\gamma \over 4}s^2)]\,(\gamma s^2+4\alpha v^2)+4\alpha v^2}\right)
\!u^2
+{\txt {x\over y}\gamma s^2v^2}\right\}=0\period
\ee
\pagebreak[3]

\noindent
The behavior of
$UV=ruv(\gamma s^2 + 4\alpha v^2)/y$
as $r\to 0$ and $r\to \infty$ is as follows:
\bea{UVasymp}
 r\to 0:\ && UV\sim \left \{
\begin{array}{llll}
r^2, & m\neq 0, & k=0, & \alpha \neq 0 \\*
r^{(2n+2)}, & m\neq 0, & k=0, & \alpha =0 \\*
r^{(2n+2k+2)}, & m\neq 0, & k>0 \\*
r^{(2n+2)}, & m=0, & k=1 \period
\end{array}\right.\non\\*
 r\to\infty:&&{\txt UV\sim \exp(-2\sqrt{\gamma}\,r)}\period
\eea
If the solution to the equations is such that $y$
\under{does not have a finite zero},
then integrating \pref{UVeq} from
$0$ to $\infty$ one obtains
\be{UVint}
\int\!2r\,dr \Big([1+{1 \over y^2}(v^2 +{\gamma \over 4}s^2)]\,
(\gamma s^2+4\alpha v^2)+4\alpha v^2 \Big ) u^2
\!+\!\int\! 2r\,dr\,{x\over y}\gamma s^2v^2=0.
\ee
As $r\to \infty$, since $z\to 0$ exponentially, $yx$ is positive.
Therefore, under the assumption that $yx$ does not change sign, the
integrands in both integrals in \pref{UVint}
are positive definite. We have
thus proven the following theorem.
\begin{theorem}
\label{xyTheorem}
Any solution of the field equations\/ $(\ref{Wphieq}-\ref{intcond})$ with
nonvanishing fields $u,v$ must be such that the product $yx$ has at least
one zero in the open interval\/ $(0,\infty)$.
\end{theorem}
We remark here that the theorem is valid also for solutions of
the equations linearized with respect to $u$ and $v$. The derivation
in this case parallels that of equations (\ref{Veq}--\ref{UVasymp})
above and leads to an equation like \pref{UVint}, except that the
$v^2$ terms are absent from the first integral.

\section{Static Solutions for W's in the Background of the
${{\bf\rm Z}}_{{\bf\rm NO}}$ Vortex.}
\setcounter{equation}{0}

 Recall that, for the \zno vortex, the electromagnetic
vector potential is zero. In terms of
our auxiliary fields $x$ and $y$, this translates into the condition
$$
\gamma x + (1 -\gamma) y = -{ 2\gamma n + m \over 2r}\period
$$
One then obtains
\bea {xno}
&&{\txt x=(\gamma-1)z - { 2\gamma n + m \over 2r}\comma}\\*
\label {yno}
&&{\txt y=\gamma z - { 2\gamma n + m \over 2r}}\comma
\eea
where $z-{n\over r}\equiv Z$ is the vector potential of the
Nielsen-Olesen vortex solution.

We shall first investigate the possibility of a perturbative solution
about the \zno vortex. This means that one should look for solutions of
the set of equations (\ref{seq}--\ref{intcond}) to lowest order in the
fields $u,v$. Since the source for $A_\varphi$ is proportional to
$j(r)$ given by equation \pref{zmod}, perturbations in the electromagnetic
field do not contribute to this order.
Perturbations in $s$ and $z$ will also be quadratic in $u,v$.
The equations for $s,x,y$ will then be the same as
for the \zno vortex.
The equations for $u,v$ are \pref{Wphieq} and \pref{Wreq},
linearized with respect to $u$ and $v$.
The boundary conditions for solutions of the linearized equations
are the same as the boundary conditions \pref{BCr=0} and \pref{BCinf}
of the exact equations.

Let us denote by \Zno the point in function space corresponding to the
Nielsen-Olesen field configuration of the \zno vortex.
We shall establish a condition on the possible solutions in the \zno
background. For this purpose we need the following lemmas.
\begin{lemma}
\label{z>0lemma}
The function $z=y-x$ of \Zno is positive definite for all
values of the parameter $\beta$.
\end{lemma}

In fact near $r=0$, $z\approx \nr$ is positive. As $r\to \infty$, $z$ goes
exponentially to zero. Then if $z$ were to change sign it would have to
go through a negative minimum. The equation for $z$ in the case of the
\zno vortex is
\be{ZNOzeq}
z''+{{z'}\over{r}}-{{z}\over{r^2}}-zs^2=0\period
\ee
At a negative minimum of $z$ this gives
\be{zineq}
z''=({{1}\over{r^2}}+s^2)z<0\period
\ee
But this is the condition for a maximum. Therefore $z$ cannot have a
negative minimum, hence it cannot change sign.
\begin{lemma}
\label{z0<0lemma}
For \Zno the value of $z_0$ in the expansion
$z=\nr +z_{0}r+...$\/ near the origin is negative.
\end{lemma}

In fact the equation for $z$ can be written
\be{zeqalt}
{{d}\over{dr}}\Big ({{1}\over{r}}{{d}\over{dr}}(rz)\Big ) - zs^2=0\period
\ee
Integrating from $0$ to $\infty$ one obtains
\be{z0int}
z_0=-\hf \int_0^\infty zs^2dr \period
\ee
Since $z$ is positive definite it follows that $z_0<0$. For
$\beta=1,\,\,z_0=-1/4.$

We are now ready to prove the following theorem.
\pagebreak[3]
\begin{theorem}
\label{mTheorem}
Any solution of the field equations\/ $(\ref{Wphieq}-\ref{Wreq})$,
with nonvanishing $u,v$ and the fields $(s,x,y)$ fixed at their
\zno vortex values, {\em or\/} these same equations linearized with
respect to $u$ and $v$, must satisfy the condition $-2n < m < 0$.
\end{theorem}

As was already pointed out, Theorem \ref{xyTheorem} holds true for
solutions of the linearized equations as well as of the exact equations.
Consider now the asymptotic expressions for
$y$ and $x$,
$$
\begin{array}{ll}
{\rm As\ }r\to 0:&
x=-{{2n+m}\over{2r}} + (\gamma-1)z_0 r +
{\cal O}(r^3)\comma\\*[2mm] ~& y=-\mr + \gamma z_0r + {\cal O}(r^3)
\comma\\*[2mm]
{\rm As\ }r\to \infty:& x=y=-{{2\gamma n+m}\over {2r}} \period
\end{array}
$$
Recall that $z_0<0$ for all values of ($\beta,\gamma$).

Assume that the bound on $m$ is not satisfied. Then we have
the following two cases:
\newcounter{mcase}
\begin{list}%
{\arabic{mcase})}%
{\usecounter{mcase}\setlength{\parsep}{0pt}\setlength{\itemsep}{0pt}}
\item $m\geq 0$. Then  $x$ and $y$ are both negative as
$r\to 0$ and at large $r$.
\item $m\leq -2n$. Then $x$ and $y$ are both positive as $r\to 0$ and
at large $r$.
\end {list}
A possible exception occurs for $m=-2n$ and $\gamma=1$.
In this case $x(r)\equiv 0$ and the condition (\ref {UVint}) for the existence
of a solution would require $u(r)\equiv 0$.
 But then the energy would always increase for any non-vanishing
configuration of $v(r)$. For this reason this case is excluded from
consideration.
In all other cases, if $x$ or $y$ were to change sign, they would
have to go through a positive maximum and a negative minimum.
But $x$ and $y$ satisfy respectively the equations
\be{ZNOyeq}
y''+{{y'}\over{r}}-\yr - \gamma zs^2=0\comma
\ee
\be{ZNOxeq}
x''+{{x'}\over{r}}-{{x}\over{r^2}} -(\gamma -1)zs^2=0\comma
\ee
where, as already shown, $z$ is positive definite.
Then, at a positive maximum of $y$, one would have
\be{yineq}
y''=\yr+\gamma zs^2>0\comma
\ee
which is the condition for a minimum. At a negative minimum of $x$
\be{xineq}
x''={{x}\over{r^2}}+(\gamma-1)zs^2<0\comma
\ee
which is the condition for a maximum.
Therefore neither $y$ nor $x$ changes sign and, by Theorem \ref{xyTheorem},
no solution exists, which proves the theorem.

Let us now apply Theorem \ref{mTheorem} to perturbative solutions about the
\zno vor\-tex. The linearized equations lead to the same constraints
\pref{C1} on $m$, $n$, and $k$ as the exact equations for all fields.
 According to these constraints, a solution of the equations
near $r=0$ is possible only for $m\geq 0$ or $m \leq -2n-1$. But Theorem
\ref{mTheorem} tells us that no global solution exists
for these values of $m$.
We thus arrive at the following result:
\begin{theorem}
\label{pertth}
In the one-component Higgs gauge,
a perturbative solution of the Euler-Lagrange equations about the \Zno
configuration does not exist for any values of
$\beta$ and $\gamma$, ($\beta>0$, $0\leq \gamma \leq 1$).
\end {theorem}
Our field ansatz is ill-defined for $\gamma=0$, but in this case
the physical $Z$ field is aligned with the U(1) hypercharge field and
does not couple at all to the W bosons.

An analysis of the variation of the
energy at \Zno, done in Ref.~\cite{JamPV} for $n=1$, shows that
the energy decreases, in a region of the parameter space
$(\beta,\gamma)$, for a perturbation in the W field with
values $m=-1$ and $k=0$.
This value of $k$ implies that the W production
is concentrated at the core of the vortex; this is natural since
there the Z field strength takes its maximal value, the Higgs
field is minimal, and the vacuum instability due to the
anomalous magnetic moment of the W boson
\cite{AmbO,MacT92,Torn93} is most pronounced.
As we have seen, for $n=1$ the values of
$m$ for which the boundary conditions at $r=0$ for the equations
(\ref{Wphieq}--\ref{intcond}) linearized with
respect to $u,v$ admit a solution, exclude the values $m=-1,-2$.
Nevertheless, within the region $(\beta,\gamma$) of instability
of the \zno vortex, one expects the energy to have a minimum for some
W configuration with $m=-1$ and the fields $s,z$ fixed at their \Zno
values. This minimum would be
a solution of the exact equations
\pref{Wphieq} and \pref{Wreq} with boundary conditions (\ref{BC2}.e) and
\pref{BCinf}.

Before proceeding with the investigation of these equations, we
shall establish the following result.

\begin{theorem}
\label{saddle}
If, for some values of $\beta,\gamma$, the equations
(\ref {Wphieq}--\ref {yeq}) with $s,z$ given by their \Zno values, admit
a solution such that the energy of the corre\-sponding state, calculated
in the quadratic approximation in the W field, is lower than that for the
\zno vortex, then for these values of $\beta$ and $\gamma$ the \zno vortex
is a saddle point in the space of field configurations.
\end{theorem}
We shall prove the theorem for the case in which $A_\varphi$
is allowed to vary and $y$
satisfies equation (\ref {yeq}) with $\alpha=\gamma$. A similar proof
can be given for the other case ($y\equiv\yno$ and $\alpha=1$).

Suppose that for some values of $\beta,\gamma$, we have a solution of the
equations (\ref{Wphieq}--\ref{yeq}) corresponding to the production of
W's and an electromagnetic potential $A_\varphi$ in the \zno background.
Let $u,v,y$ be the functions corresponding to this solution and write
$y=\yno+\ybar$, $x=\xno+\ybar$, where $\xno,\yno$ are the values of
$x,y$ in
the \Zno configuration, given by (\ref{xno}, \ref{yno}).
Let $\eno$ be the energy for the \zno vortex
(with $\Phi_0^2=1$ for simplicity) and let us break up the energy
difference $\delta {\cal E}={\cal E}-\eno$ into three terms:
\be {E1}
\ea\!=\!{2\pi
\over \gamma}\!\!\int\! {rdr\,[(v'+{v\over r}+2\yno u)^2
-4(y'_{{\rm\teeny NO}}+{\yno\over r})uv +\gamma (u^2+v^2)s^2]}
\comma\hfill\\*
\ee
\be {E2}
\eb\!=\!{2\pi\over \gamma}\!\!\int\! {rdr\, [4(v'\ybar -\ybar'v)u
+4(2\yno\ybar+v^2)u^2+{1\over {1-\gamma}}(\ybar'+{\ybar\over r})^2]}
\comma\hfill \\*
\ee
\be {E3}
\ec\!=\!{2\pi\over \gamma}\!\!\int\! {rdr\, (2\ybar u)^2}
\period
\ee
Since $\ybar$ is already second order in ($u,v$), then
$\ea$ is the lowest order energy shift which is
assumed to be negative, $\ea<0$.

Consider configurations of the fields $\ybar_{\lambda}=
\lambda \ybar,u_{\lambda}=\sqrt{\lambda}u,
v_{\lambda}=\sqrt{\lambda}v$, where $\lambda$ is a scaling factor.
The energy corresponding to these configurations with $s,z$ at their \Zno
values will be given by
\be {ET}
{\cal E}(\lambda)=\eno+\lambda \ea+\lambda^2\eb+\lambda^3 \ec \period
\ee
This is a cubic polynomial in $\lambda$ with extrema at
\be {lambda}
\lambda_{\pm}={1\over 3\ec}{\Big (}-\eb \pm \sqrt{\eb^2-3\ec\ea}\,\Big )
\period
\ee
Since $\ea<0$, and $\ec>0$, only $\lambda_{+}$ is a positive
root corresponding to a minimum of ${\cal E}(\lambda)$.
Thus in the interval $0<\lambda<\lambda_{+}$ the energy
${\cal E}(\lambda)$ will be smaller than ${\cal E}(0)=\eno$.
Therefore there will be configurations of arbitrarily small fields
$\ybar_{\lambda},u_{\lambda},v_{\lambda}$ for which the
energy becomes smaller than $\eno$. This proves the theorem.

Now, by construction, ${\cal E}(\lambda)$ has an extremum at $\lambda=1$.
Therefore the value of $\lambda$ at the minimum of ${\cal E}(\lambda)$
is $\lambda_{+}=1$.
Inserting this value in the expression (\ref {lambda}),
one obtains
\be {EEE}
\ea=-(2\eb+3\ec) \period
\ee
This is a very useful result for numerical computations since it provides
a good check on the precision of the calculation of the energy shift
$\delta{\cal E}$,
\be {delE}
\delta {\cal E}\equiv
{\cal E}_1+{\cal E}_2+{\cal E}_3={1\over 2}(\ea-\ec)=-(\eb+2\ec) \period
\ee
\section{Numerical Solutions}
\setcounter{equation}{0}
In this section we present solutions of the set of equations
(\ref {Wphieq}--\ref {Wreq}) for W's in the presence of the fields
$s,x,y$ given by their
\Zno values.
The second-order equation \pref{Wphieq} can be written in the form of two
first-order equations for the functions $rv$ and
$F\equiv v'+{v\over r}+2uy$, as
\bea{rveq}
(rv)'&=&r(F-2uy)\comma\\*
\label{Feq}
F'&=&(\gamma s^2+4u^2)v-2(y'+{y\over r})u\period
\eea
Here $y=\gamma Z-{m\over 2r}$, where $Z$ and $s$ are given by the
Nielsen-Olesen vortex solution and $m=-1$.
Equation \pref{Wreq} was used to solve algebraically for $u$ in
terms of $v$ and $F$,
\be{usolve}
u=2{(y'+{y\over r})v-yF\over \gamma s^2+4v^2}\period
\ee
\renewc{\thefootnote}{\fnsymbol{footnote}}
\begin{table}
\begin{center}
\parbox{3.5in}{
$$
\begin{array}{|r|l|l|r|l|}\hline\hline
\multicolumn{1}{|c|}{\beta} &
\multicolumn{1}{c|}{\gamma} &
\multicolumn{1}{c|}{v_0} &
\multicolumn{1}{c|}{v_1} &
\multicolumn{1}{c|}{\delta{\cal E}}\\ \hline
.5  &  .25 &  -.092750  &  .06860281  &  -.00028\\
.5  &  .5  &  -.211671  &  .41657161  &  -.0106\\
1   &  .5  &  -.237408  &  .35632791  &  -.0133\\
.5  &  1   &  -.377157  & 1.04195850  &  -.0846\\
1   &  1   &  -.418631  &  .79309235  &  -.1003\\
\hline\hline
\end{array}
$$
\caption{\label{bgtable}{
Boundary values and change of energy due to W-boson condensation
in the fixed background
of the \zno vortex, for a selection of parameters
$\beta$ and $\gamma$.}}
}
\end{center}
\end{table}

\vspace*{-7mm}
The system of equations was first propagated
from $r_1=18.5$ down to \mbox{$r_0=0$}. The boundary
condition $v_0=-u_0$ imposed at this point allowed for a precise
determination of $v_1$, while $v_0$ was determined with less precision.
 Values of these two parameters, which determine the boundary conditions
at $r=0$ and at $r=\infty$, are listed in Table \ref{bgtable}
for several different values of $\beta$ and $\gamma$.
 Next, using these boundary values, the  equations were
propagated from $r_1$ and from $r_0=0$ to a matching point in the interval
$(1.5,2.5)$. The values of $u$ and $v$ agreed within a precision of $10^{-2}$
at matching points in this interval. Both $u$ and $v$ vary
monotonically without changing sign.

Finally, since this is a non-perturbative solution, one has to check whether
the energy of these states with W's is actually lower than that of the
\zno vor\-tex. We computed the energy difference between each of these states
and the \zno vortex and found in each case $\delta {\cal E}<0$.
Hence in the sector $(\beta>.5$, $\gamma>.25)$ we find
that the \zno vortex is unstable against W production. The results for
$\delta {\cal E}$ in units of $\Phi_0^2$ (energy per unit length of vortex)
are also given in Table \ref{bgtable}. The precision given for the parameters
$v_0,v_1$, is that required to obtain the precision in the energy shift
$\delta {\cal E}$. However the accuracy of these numbers is sensitive to the
accuracy in the determination of the input functions $s,z$ as solutions of
the \zno vortex.

In each of the cases reported here the relations (\ref {delE})
(with $\ybar=0$, ${\cal E}_2=0$) were satisfied
within the accuracy obtained for $\delta {\cal E}$.

The line in parameter space along which the equations considered in this
section cease to have a solution with ${\cal E}_0<0$, provides an upper
bound on the boundary line $\Gamma$ for the stability region of the \zno
vortex. An upper bound for this curve has also been obtained by
Klink\-hamer and Olesen \cite{KliO94} using a different approach.

\section{Field Configuration of Minimal Energy with W
Fields}

Since the \zno vortex with $n=1$ is unstable in a region of the parameter
space $(\beta,\gamma)$ with respect to W production in a state of angular
momentum $m=-1$, and because the energy is bounded from below, one expects
that there exists some configuration of the fields, with
W's in such a state, for which the energy is a minimum.
The Euler-Lagrange equations (\ref{seq}--\ref{intcond}), however,
do not admit a solution for $m=-1$ with the
boundary conditions required to make the energy finite. This seems to
preclude the existence of a stationary state with W's at least defined
in the space of differentiable functions. A minimum energy state may
nevertheless exist in the sense of a distribution.

In order to investigate this possibility we propose to study a model
obtained from Weinberg-Salam (WS) model with the following modifications:

\vspace*{1ex}
i) Set ${\bf W}=\sqrt{\epsilon}\,{\bf W'}$

ii) Add to the energy density a term
$${1\over\gamma}\epsilon\,(1-\epsilon)
( i{\bf W'}^\dagger\times{\bf W'})^2$$
This term is positive definite for $0<\epsilon<1$. In the new model the
boundary conditions at $r=0$ can be satisfied for $m=-1$.

The model coincides with WS for $\epsilon=1$. In the instability
region of parameter space, the Euler-Lagrange equations for the fields have
a solution which, as $\epsilon \to 0$, approaches that found in the WS
model for W's in the \zno vortex background.
As $\epsilon$ increases from
$0$ to 1, either of the following possibilities may occur:
\newcounter{epscase}
\begin{list}%
{\arabic{epscase})}%
{\usecounter{epscase}\setlength{\parsep}{0pt}\setlength{\itemsep}{0pt}}
\item The equations do not have a solution for
$\epsilon>\epsilon_{{\rm max}}<1$.
\item As $\epsilon \to 1$ the solutions approach almost everywhere the
configuration in the vacuum state
and the energy approaches zero.
\item As $\epsilon \to 1$ one obtains a sequence of solutions which has no
limit but the energy has a definite limit greater than zero.
\end{list}
In cases 1) and 2) a stable vortex with W's and a finite energy does not
exist. In case 3) the existence of a stable vortex depends on establishing
the stability of the solutions as $\epsilon \to 1$.

\section*{Feza G\"{u}rsey in Memoriam}
At this conference one of us (S.W.M.) presented
a special contribution in memory
of Feza G\"{u}rsey, which will be published elsewhere
in these proceedings.
The other (O.T.) wishes to express in this space his feelings of
gratitude to Feza G\"{u}rsey for his inspired teaching of
quantum field theory at Yale University and for the many ways
graduate students have benefitted, through the years,
from his vibrant personality.


\begin{thebibliography}{99}
\setlength{\parsep}{-1mm}
\setlength{\itemsep}{0pt}

\bibitem{NieO73}
H. B. Nielsen and P. Olesen, \npb{61}{45}{73}.
\bibitem{Nam77}
Y. Nambu, \npb{130}{505}{77}.
\bibitem{Vac92}
T. Vachaspati, \prl{68}{1977}{92}; \ibid{69}{216}{92}.
\bibitem{Wein67}
S. Weinberg, \prl{19}{1264}{67}.
\bibitem{Sal68}
A. Salam in {\em Elementary Particle Theory\/},
edited by N. Svartholm (Almqvists F\"{o}rlag AB, Stockholm, 1968) p. 367.
\bibitem{AmbO}
J. Ambj{\o}rn and P. Olesen, \npb{315}{606}{89};\
\npb{330}{193}{90};\
\ijmpa{5}{4525}{90}.
\bibitem{MacT92}
S. W. MacDowell and O. T\"{o}rnkvist,
\prd{45}{3833}{92}.
\bibitem{Torn93}
Nils Ola T\"ornkvist, Yale University Ph. D. Thesis (Nov. 1993) RX-1493,
Microfiche UMI-94-15879-mc.
\bibitem{JamPV}
M. James, L. Perivolaropoulos, and T. Vachaspati, \prd{46}{R5232}{92};\
\npb{395}{534}{93}.
\bibitem{Perk93}
W. B. Perkins, \prd{47}{R5224}{93}.
\bibitem{AchGHK93}
A. Ach\'{u}carro, R. Gregory, J. A. Harvey, and K. Kuijken,
\prl{72}{3646}{94}.
\bibitem{Torn92}
O. T\"ornkvist, Yale Preprint
YCTP-P11-92 (Apr. 1992),\\
HEP-PH/9204235.
\bibitem{Oles93}
P. Olesen, Niels Bohr Institute preprint NBI-HE-93-58
(Oct. 1993),\\
HEP-PH/9310275.
\bibitem{Peri93}
L. Perivolaropoulos, \prd{48}{5961}{93};\ {\em ibid.\/} note to appear.
\bibitem{KliO94}
F. R. Klinkhamer and P. Olesen, \npb{422}{227}{94}.
\end{thebibliography}
\end{document}